# DEPLETION LAYER-INDUCED SIZE-EFFECTS IN FERROELECTRIC THIN FILMS: A GINZBURG-LANDAU MODEL STUDY

Nathaniel Ng, Rajeev Ahluwalia, David J. Srolovitz

*Institute of High Performance Computing, Agency for Science, Technology and Research*

*1 Fusionopolis Way, #16-16 Connexis, Singapore 138632*

**Abstract**

A Ginzburg-Landau model is used to demonstrate how depletion layers give rise to thickness-dependent ferroelectric properties in thin films. It is shown that free charge layers at the film-electrode interface can result in an internal electric field in the bulk of the film even when no external voltage is applied. At high values of the donor dopant density and small thicknesses, this internal electric field can be strong enough to lead to the formation of a domain pattern. This causes a drop in the remnant polarization; a direct demonstration of the important role free charge plays in thin ferroelectric films.

## 1. Introduction

The progressive miniaturization of devices based upon ferroelectric phenomena has led to an increased interest in size effects in ferroelectric thin films [1-2]. An important issue for device applications is the question of whether a remnant state (i.e., a macroscopically polarized state in the absence of an external electric field) exists within the ferroelectric film as the film thickness is decreased. Such size effects are strongly influenced by the electrical boundary conditions that exist at the ferroelectric-electrode interface [2]. A large body of research on the role of uncompensated bound charges at the electrode [3-4] and associated depolarization fields exists [5-8], including first-principles work [9-10]. Theory suggests that because of depolarization effects, two critical length scales exist, one below which a single domain remnant state splits into a multi-domain state with 180° domains and the second where ferroelectricity completely disappears [7, 11]. However, these analyses of size effects assumed that the ferroelectric is an electrical insulator. In reality, ferroelectrics are usually wide band gap semiconductors; this fact can lead to the formation of free charge layers at the ferroelectric–electrode interface. These charge-depleted layers arise from the differences in the work functions between the metal and the ferroelectric [2, 12], and cause migration of free charges from the ferroelectric into the metal. Depletion layers may also form due to the accumulation of oxygen vacancies at the ferroelectric-electrode interface after repeated cycling of the electric field [12]. Thus, in addition to the traditionally studied bound charge effects, free charge can strongly influence the electrostatic boundary conditions in thin film devices. How do these charge depleted layers influence ferroelectric behavior? This question becomes of central importance for thin films where the film thickness becomes comparable to the depletion layer thickness. An understanding of this issue is key, not only from a fundamental standpoint, but for its important implications for device applications.

The role of depletion layers in ferroelectric thin films has received some attention in literature. Xiao *et al.* [13] demonstrated the formation of a depletion layer based on a Ginzburg-Landau model that incorporated mobile charges. Bratkovsky and Levanyuk [14] showed that depletion charge near the electrode also reduces the ferro-/para-electric transition temperature. Baudry and Tournier [15] and Zubko *et al.* [16] studied the influence of depletion layers on the remnant polarization and hysteresis loops within the Landau theory framework. An analytical study based on a linear approximation to this problem was performed by Tagantsev *et al.* [17]. Recently, more sophisticated treatments of the influence of depletion layers on ferroelectrics have been made [18-20]. However, the full implications of the effects of depletion layers on size effects in thin, ferroelectric films remain unexplored. Specifically, the following important issues have not received adequate attention. What is the influence of the depletion layers on the domain patterns as the film thickness is decreased? Do depletion layers have any effect on domain nucleation and growth during switching? In this paper, we employ Ginzburg-Landau theory and phase field simulations to demonstrate that depletion layers have very important consequences for size effects in ferroelectric thin films.

The paper is organized as follows. In section 2, we describe the Ginzburg-Landau model for a thin film with depletion layers as well as the kinetic model used to simulate the polarization dynamics. Section 3 describes the results on the stability of the remnant monodomain states under short-circuit boundary conditions. We analyze these results via a 1D analytical model in section 4. The role played by depletion layers on polarization switching is discussed in section 5. Lastly, we end the paper with a summary and conclusion.

**2. Ginzburg-Landau model**

In the Ginzburg-Landau-Devonshire framework, the total free energy of the system,

$F_T$, is:

$$F_T = \int d\vec{r}\left[f_L + f_G + f_{el}\right] \qquad (1)$$

where $f_L$, $f_G$, and $f_{el}$ represent the local Landau, gradient and elastic energy densities, respectively. The Landau energy is obtained from the symmetry-allowed expansion of the free energy as an eighth order polynomial in the polarization components of $\vec{P}$. Substrate effects are accounted for by assuming a homogeneous in-plane biaxial strain ε that arises from the lattice mismatch between film and substrate. Since our focus is the role of depletion layer induced electrostatic effects, we restrict our study to thin films in the region of the phase diagram where they are effectively uniaxial ferroelectrics. This is achieved when the compressive misfit strain is sufficiently large to suppress in-plane ferroelectric polarization [21]. We consider an effective free energy that is nonlinear in $P_z$ (polarization normal to the film); to eighth order in Landau theory [22], this is expressed as

$$f_L + f_{el} = \tfrac{1}{2}\alpha_1^*\left(P_x^2 + P_y^2\right) + \tfrac{1}{2}\alpha_3^* P_z^2 + \tfrac{1}{4}\alpha_{11}^* P_z^4 + \tfrac{1}{6}\alpha_{111} P_z^6 + \tfrac{1}{8}\alpha_{1111} P_z^8 \qquad (2)$$

where $\alpha_1^*$, $\alpha_3^*$, $\alpha_{11}^*$, $\alpha_{111}$, $\alpha_{1111}$ are appropriate material constants [21]. The appropriate contribution from the polarization gradients at domain walls is expressed as

$$f_G = \tfrac{1}{2} K\left[\left|\vec{\nabla} P_x\right|^2 + \left|\vec{\nabla} P_y\right|^2 + \left|\vec{\nabla} P_z\right|^2\right]. \qquad (3)$$

The electric field, **E**, is obtained by solving Gauss's Law

$$\nabla \cdot \mathbf{D} = \nabla \cdot (-\varepsilon_0 \nabla \phi + \mathbf{P}) = \rho, \qquad (4)$$

where

$$\rho(z) = \begin{cases} \rho_0, & z < w \text{ or } d - z < w \\ 0, & \text{otherwise} \end{cases}, \qquad (5)$$

the film thickness is $d$, the depletion layer thickness is $w$, and $\rho_0 = qN_D$ and $N_D$ are the space charge and donor dopant density. As is well known, the depletion layer thickness depends on the dopant density and the applied and built-in potentials. In the present work, we describe the depletion layer width dependence on the dopant density as $w = \sqrt{2V_{bi}\varepsilon/N_D q} = C/\sqrt{N_D}$ [2, 13, 16-17, 23-24], which has been successfully used to model the semiconducting nature of ferroelectrics in recent works [16-17, 24]. Here $V_{bi}$ is the built-in potential, ε is the dielectric constant, and $C = \sqrt{2V_{bi}\varepsilon/q}$ is a constant that represents a property of the ferroelectric-electrode interface. We should remark that while there are more sophisticated calculations which obtain the depletion layer widths in a self consistent manner [18], we use the current expression in the present work to simplify the analysis.

We choose parameters appropriate for a BaTiO$_3$ thin film on a SrRuO$_3$/SrTiO$_3$ substrate with a compressive misfit strain $\varepsilon = -0.022$ [7] and gradient coefficient $K = |\alpha_1|\delta^2$, where $\delta$ is the smallest length scale resolved in the simulation, which is taken to be $\delta = 1$ nm in the present simulations. Note that we have not included a background dielectric permittivity of the ferroelectric in equation (4) [25]. This common assumption [13, 15-16] is reasonable here since the background dielectric constant is usually much smaller than the dielectric constant of the ferroelectric. Although there are situations in which inclusion of the background dielectric permittivity may be important, it has little effect on the main conclusions presented here.

Since we are interested in simulating domain patterns and polarization switching, a model for the dynamics of the polarization fields is essential. The polarization kinetics is studied within the time-dependent Ginzburg-Landau framework:

$$\frac{\partial P_i}{\partial t} = -\Gamma \left[ \frac{\delta F_T}{\delta P_i} - E_i \right] \qquad (6)$$

Here, $\Gamma$ is a kinetic coefficient related to the domain wall mobility and $E_i$ is the component of the electric field, evaluated using equations (4) and (5). We use these equations to test the stability of a single domain remnant state as a function of the donor dopant density $N_D$ and the thickness $d$ and also investigate the effect of the depletion layers on polarization switching. We use natural boundary conditions $\partial P_z/\partial z = 0$ for the polarizations at the interfaces at $z = 0$ and $z = d$, which means that the interface energy does not depend on the polarization. Further, we also assume that a surface charge exactly compensates the bound surface charge due to the polarization discontinuity at the ferroelectric-electrode interfaces. Equation (6) is discretized using finite differences. The lengths are measured in the units of the smallest length scale $\delta$ and a scaled time step of $\Delta t' = \Gamma |\alpha_1| \Delta t$.

## 3. Stability of the Monodomain Remnant State

We first address the technologically important question: How stable is a single domain remnant state when depletion layers with free charge are present? To address this issue, we initialize the entire film in a mono-domain state by setting the polarization at each point in space to $P_x = P_y = 0$ and $P_z = P_s$ ($P_s$ is the spontaneous polarization) plus an initial noise chosen at random from the interval $\pm P_i$, where $P_i = 0.001\, P_s$. The TDGL equations are integrated with short-circuit boundary conditions to study the stability of this remnant state.

As discussed earlier, the depletion layer width $w$ is described by $w = C/\sqrt{N_D}$ [17,23], where $C$ is a property of the ferroelectric-electrode interface that depends on the built in potential and the dielectric constant. In the present simulations, we set $C = 1.02 \times 10^5 \, \text{m}^{-0.5}$ which corresponds to a depletion layer width of $w = 16$ nm for a dopant density $N_D = 4.11 \times 10^{25}$ m$^{-3}$. We confirmed that choosing larger value of $C$ did not qualitatively change the nature of the mono- to multi-domain transitions shown in Fig. 1. We note, however, that the value of $N_D$ at which the transition occurs does depend on the value of $C$. This effect is captured by the analytical solutions shown below. The TDGL simulations are performed for a series of $N_D$ values and film thicknesses $d$.

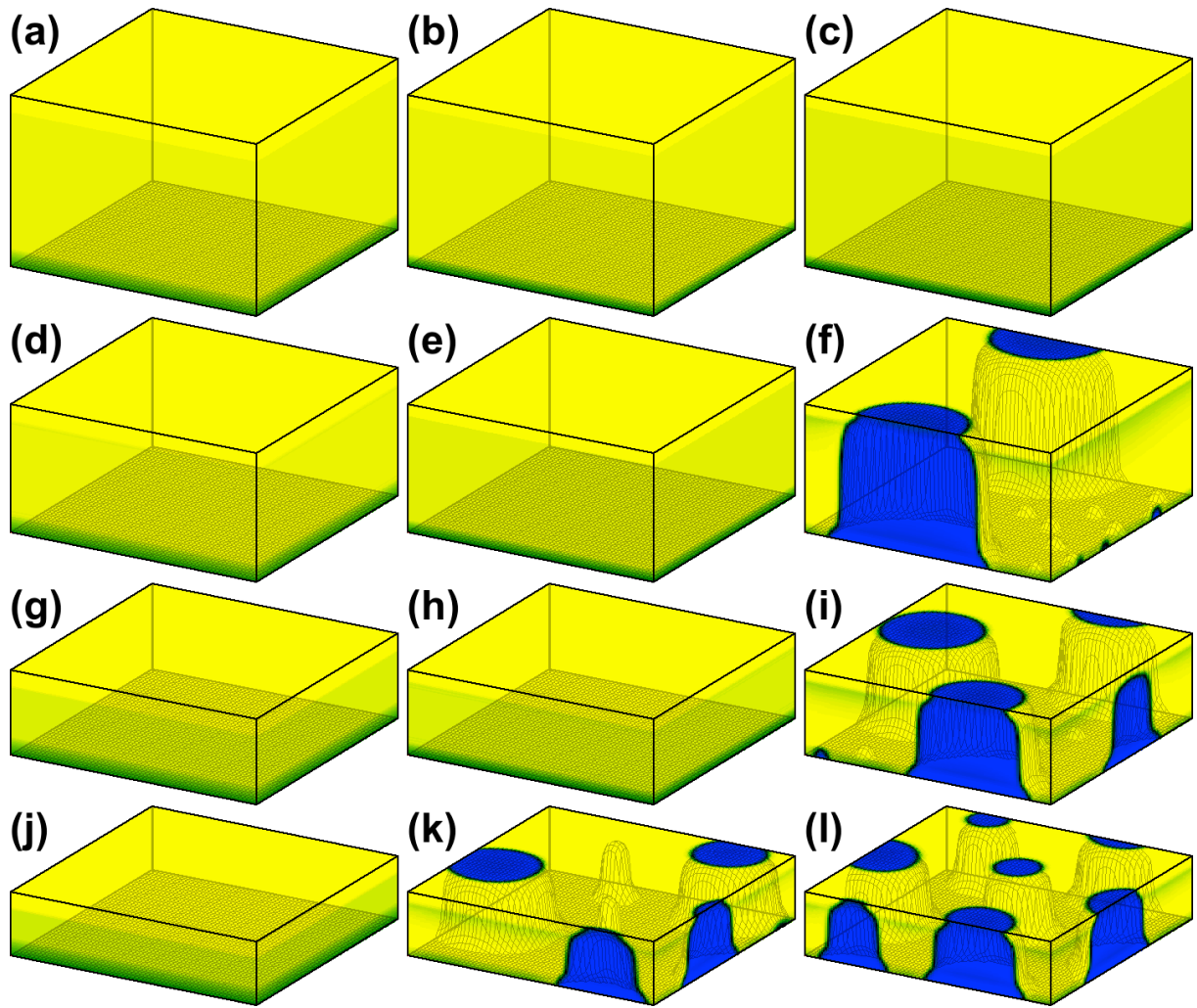

Figure 1 (color online): Final equilibrated domain structures from the TDGL simulation as a function of thickness (vertical axis – $d$ = 48, 64, 96, 128 grid points) and donor dopant density (horizontal axis - $N_D$ = 5.48, 9.59, and 11.0 × 10$^{25}$ m$^{-3}$). The depletion layer width is chosen as $w = CN_D^{-1/2}$, rounded to the nearest grid point.

At small donor dopant densities ($N_D \leq 5.48 \times 10^{25}$ m$^{-3}$), the initial single domain state is found to be stable for all thicknesses (Fig. 1 a, d, g, j). However, for $N_D = 9.59 \times 10^{25}$ m$^{-3}$, the single domain state is stable for large thicknesses (Fig. 1 b, e, h) and unstable at small thicknesses (see Fig. 1k for $d \leq 48$nm where the film splits into a multi-domain state with the formation of reverse polarization domains). At higher values of $N_D$ ($N_D \geq 11.0 \times 10^{25}$ m$^{-3}$), the film splits into domains at larger thicknesses ($d \leq 96$nm) (Fig. 1 f, i, l). This shows that the mono-domain to multi-domain transition predicted by the present simulations can become crucial when the dopant densities are high. The appearance of this domain pattern may have important consequences for remnant polarization in ferroelectric thin films as it implies a reduction in the remnant polarization. While these results show that the remnant state is unstable with respect to the formation of reverse-domains in the presence of free charge, a similar splitting of the remnant state into striped 180° domains has been predicted for films with uncompensated bound charge at electrodes [7]. However, unlike in the bound charge case, the transition observed here is not sharp and the domain formation is localized. Depending on the value of $N_D$, domain formation due to depletion layer effects can be seen for thicknesses as large as ~13 times the width of the depletion layer (see Fig. 1c).

It is interesting to analyze the shape of the domains observed in Fig. 1. There are two distinct kinds of domains: (i) cylindrical domains that stretch from the bottom to top interface - expanding (flaring) into the bottom depletion layer and tapering in the top layer (Fig. 1 f, i, l) and (ii) domains that stretch from the bottom interface and terminating within the bulk of the film (Fig. 1 f, i, k). These domains take these shapes in order to minimize the electrostatic energy associated with charged domain walls (polarization vectors that are head-to-head or tail-to-tail). The formation of these shapes lead to rotated polarization vectors in the interfacial regions to avoid charged domain walls.

What causes the mono- to multi-domain transition? To understand this, we examined the distribution of electric fields in the film and found that a non-zero electric field, opposite to the polarization direction, exists in the bulk of the film where there is no free charge. Moreover, the magnitude of this internal electric field increases with increasing $N_D$ and decreasing film thickness. The existence of this internal electric field and its dependence on the film thickness and the dopant density is explicitly shown in the analytical solution presented in section 4. For moderate values of $N_D$, the internal electric field everywhere in the film is lower than the thermodynamic coercive field ($E_z > -E_c$). No domain formation occurs for these cases. However, as we consider films with higher $N_D$, the electric field in the bottom depletion layer approaches the thermodynamic coercive field. At a finite value of $N_D$, the magnitude of the electric field inside the layer exceeds the thermodynamic coercive field and domains are nucleated within the bottom depletion layers. This situation is schematically depicted in Fig. 2. Note that at the tip of these domains (for example at point A in Fig. 2), there will be an electric field concentration (see Fig. 7) due to the depolarization fields. As $N_D$ is increased further, the electric field concentration at the tip becomes larger than the thermodynamic coercive field (although the average electric field in the bulk regions may still be smaller). When this happens, a domain will nucleate in the bulk regions, leading to the formation of a multi-domain state, similar to those in Fig. 1 f, i, k and l.

At this stage, some remarks on the experimental validity of this mono-domain to multi-domain transition are in order. The internal electric fields in perovskites for typical doping densities are much lower than the thermodynamic coercive fields. The mono- to multi-domain transition will be important when either the dopant density is very high and/or for materials for which the thermodynamic coercive field is low (e.g., at temperatures close to the second order transition points). Even for the cases where the built-in electric field is low, the presence of defects may still lead to the nucleation of reverse domains.

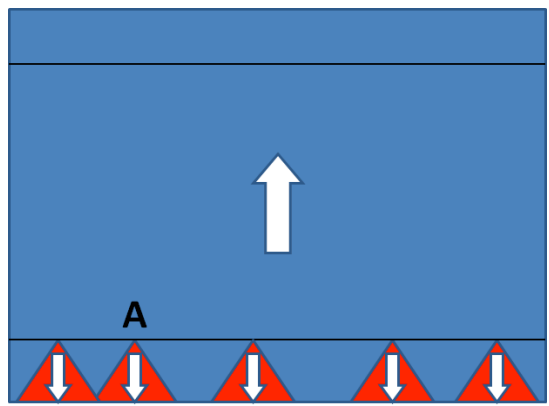

Figure 2 (color online): 2D schematic depicting the appearance of reversed domains inside the bottom depletion layer. Electric field concentrations at the tip of these domains (Fig. 9) can cause nucleation of domains inside the bulk of the film where there is no depletion charge.

## 4. One Dimensional Analytical Solution

The basic physics associated with 3D simulation results can be deduced from a simplified one-dimensional model. An analytical solution for the electric field distribution inside films with depletion layers was obtained by Tagantsev et al. [17] for the special case of a linear dielectric material. Here we derive an analytical solution that takes into account the non-linearity of the dielectric constant – an important feature of ferroelectrics. We start with the Landau free energy (to fourth order in the polarization):

$$F_T = \tfrac{1}{2}\alpha_1 P_z^2 + \tfrac{1}{4}\alpha_{11} P_z^4 \qquad (7)$$

The appropriate electric field

$$E_z = \frac{\partial F_T}{\partial P_z} = \alpha_1 P_z + \alpha_{11} P_z^3, \qquad (8)$$

Next, we rewrite Gauss's Law as: $\varepsilon_0 \frac{\partial E_z}{\partial P_z}\frac{\partial P_z}{\partial z} + \frac{\partial P_z}{\partial z} = \rho(z)$, or $\frac{\partial P_z}{\partial z} = \rho(z)\left[1 + \varepsilon_0 \frac{\partial E_z}{\partial P_z}\right]^{-1}$. Since $\chi$ is large in ferroelectrics, $1/\chi = \varepsilon_0(\partial E_z/\partial P_z) = \varepsilon_0\left[\alpha_1 + 3\alpha_{11} P_z^2\right]$ is small. Taylor-expanding to first order in $\varepsilon_0(\partial E_z/\partial P_z)$ we can write:

$$\frac{\partial P_z}{\partial z} = \rho(z)\left[1 - \varepsilon_0 \frac{\partial E_z}{\partial P_z}\right] \qquad (9)$$

Integrating equation (9) in the depletion layers and setting $P_z = P_b$ for $w < z < d - w$, we find

$$P_z = \begin{cases} a\tanh\{b\rho_0(z-w)\} + P_b, & z < w \\ P_b, & w < z < d-w, \\ a\tanh\{b\rho_0[z-(d-w)]\} + P_b, & z > d-w \end{cases} \qquad (10)$$

where $a = \sqrt{(1-\alpha_1\varepsilon_0)/(3\alpha_{11}\varepsilon_0)}$, $b = \sqrt{3\alpha_{11}\varepsilon_0(1-\alpha_1\varepsilon_0)}$, and $P_b$ represents the polarization in the bulk of the film where there is no free charge. $P_b$ can be computed by applying the boundary condition, $-U = \int_0^d E_z dz$, where $-U/d = E_{ext}$ is the contribution from the external electric field. Using (8) and (10) we obtain the renormalized equations from which the effective hysteresis loop for the film can be expressed as

$$E_{ext} = \tilde{\alpha}_1 P_b + \alpha_{11} P_b^3, \qquad (11)$$

where $\tilde{\alpha}_1 = \left\{\alpha_1 + 2\frac{w}{d}\left(\frac{1}{\varepsilon_0} - \alpha_1\right)\left[1 - (b\rho_0 w)^{-1}\tanh(b\rho_0 w)\right]\right\} \qquad (12)$

The combined effects of space charge and thickness is to simply shift the effective value of $\alpha_1$. Here, $P_b$ can be obtained using the standard solution for a cubic equation. Under zero external field, $P_b = \pm\sqrt{-\tilde{\alpha}_1/\alpha_{11}}$, i.e.,

$$P_b/P_s = r = \pm\sqrt{1-2\frac{w}{d}\left[1-(\alpha_1\varepsilon_0)^{-1}\right]\left[1-(b\rho_0 w)^{-1}\tanh(b\rho_0 w)\right]} \qquad (13)$$

where $P_s = \sqrt{-\alpha_1/\alpha_{11}}$ is the spontaneous polarization for the film without depletion layers. The corresponding bulk electric field, $E_b$, is calculated using $E_b = \alpha_1 P_b + \alpha_{11} P_b^3$ as:

$$E_b/E_c = \mp \tfrac{3\sqrt{3}}{2} r(1-r^2) \qquad (14)$$

where $E_c$ is the intrinsic (thermodynamic) coercive field of the film without depletion layers.

Combining equations (10) and (13), we obtain a closed form expression for the polarization profile, $P_z(z)$, that includes the effects of both space charge and thickness dependence. Substitution into equation (8), $E_z = \alpha_1 P_z + \alpha_{11} P_z^3$, yields an explicit formula for the electric field profile.

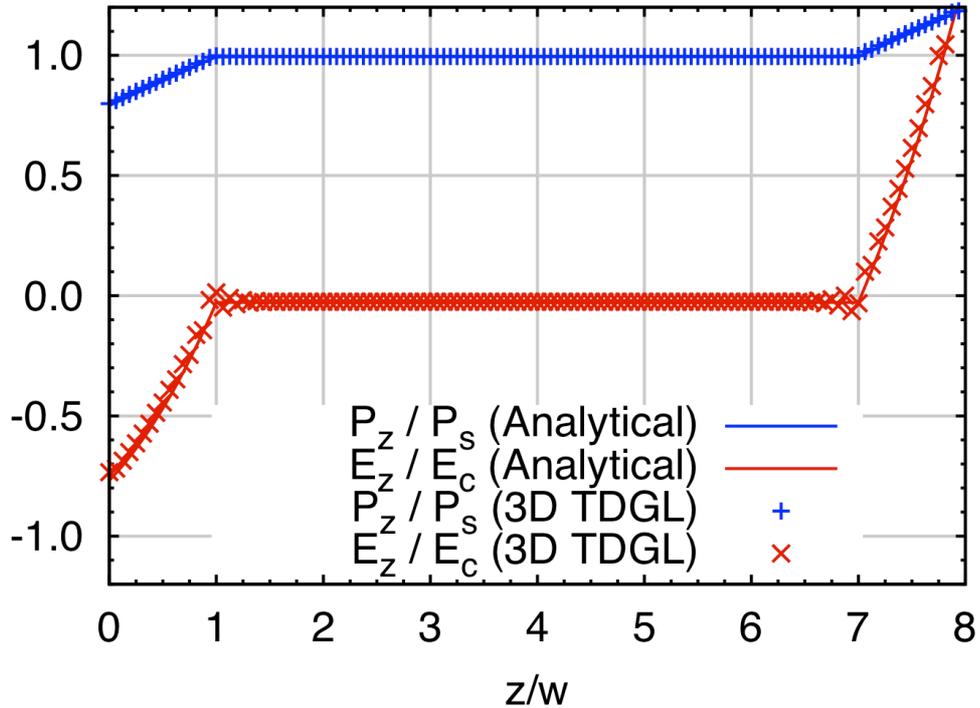

Figure 3 (color online): Polarization and Electric Field Profiles: $P_z / P_s$ and $E_z / E_c$ vs $z$ ($N_D = 2.74 \times 10^{25}$ m$^{-3}$) obtained from the analytical solution as compared to the 3D simulation with the same 4th order Landau parameters

The profiles of $P_z$ and $E_z$ have been validated against the 3D TDGL simulations as shown in Fig. 3. This agreement holds even though the analytical solution ignores the value of the gradient coefficient, $K$. Since the length scale of our simulation is set by our choice: $K' = K/(|\alpha|\delta^2) = 1.0$, where $K'$ is inversely proportional to $\delta^2$, it is believed that this agreement will hold even if the length scale $\delta = 1$nm is changed, meaning that our results should hold if we rescale our simulation (by changing $\delta$) within reasonable limits. Note also that since the analytical solution has been obtained using 4$^{th}$ order Landau theory, we have appropriately chosen the parameters such that the equilibrium polarization is the same as the full 8$^{th}$ order expansion considered in the 3D simulations of Fig. 1. For typical values of $N_D$ (such as those used by Zubko *et al* [16]), we can expand the hyperbolic tangent term in Eq. (13) in a Taylor series. Truncating at 4th order gives:

$$\frac{P_b}{P_s} = r \cong \pm\sqrt{1 - \frac{2}{3}\frac{b^2 w^3 q^2 N_D^2}{d}\left[1-(\alpha_1\varepsilon_0)^{-1}\right]} \quad (15)$$

From this analysis, we see that a non-zero electric field $E_b$ exists (except in the $w/d \to 0$ or $\rho_0 \to 0$ limits where $|P_b| = P_s$) even for short-circuit boundary conditions ($E_{ext} = 0$). The non-zero $E_b$ explains the internal electric field observed in the 3D simulations (Fig. 3). Examination of Eqs. (14) and (15) suggests that it is useful to introduce a dimensionless distance:

$$d' = \frac{d}{b^2 w^3 q^2 N_D^2\left[1-(\alpha_1\varepsilon_0)^{-1}\right]} \quad (16)$$

By substituting $w = C/\sqrt{N_D}$ into Eq. (16), we see that the key parameter that determines the scaling of the bulk polarization $P_b/P_s$ or the bulk electric field $E_b/E_c$ is $d/N_D^{1/2}$ or $dw$.

How does the internal electric field, $E_b$, vary with the film thickness and the dopant density? To address this, we plot (see Fig. 4) $E_b$ versus the rescaled thickness, $d'$, obtained from the analytical solution (Eqs. 13 and 15). Clearly, the magnitude of the internal electric field increases with decreasing $d'$ until $d' = 1$, where the bulk electric field, $E_b$, becomes equal to the intrinsic coercive field. In a full 3D simulation, this large internal field in the bulk and depletion layers leads to the mono-domain to multi-domain transition that was observed in Fig. 1 (f, i, k, and l). Note that the film actually splits into a multi domain state well before $|E_b/E_c| = 1$; as $|E_z/E_c|$ at $z = 0$ reaches the coercive field for switching earlier than $|E_b/E_c|$ (see Fig. 3).

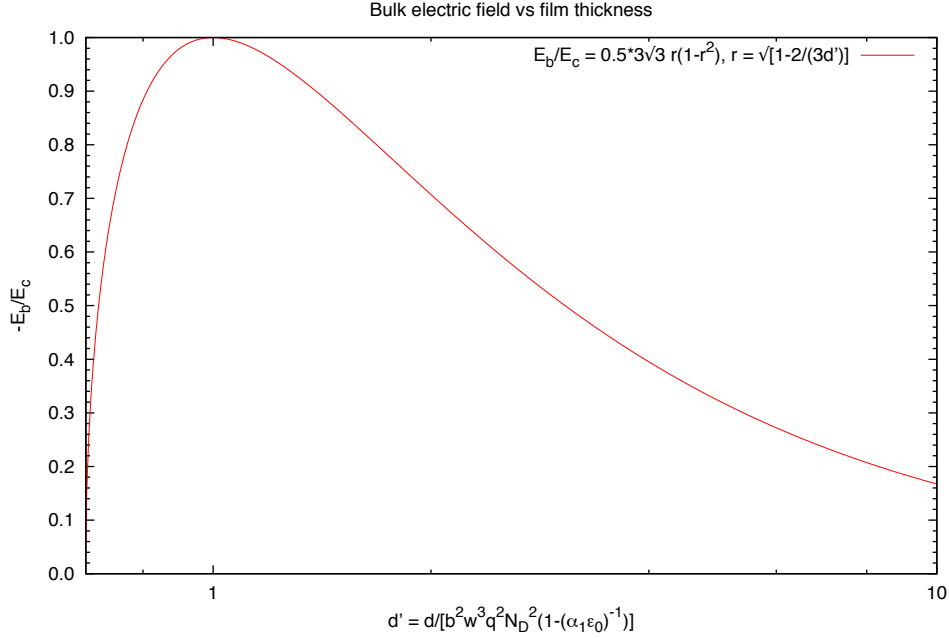

Figure 4: Rescaled internal electric field, $-E_b / E_c$, vs rescaled film thickness $d'$.

We can draw useful conclusions about the thickness dependence and dopant density dependence of the internal electric field by examining Fig. 4. Figure 4 clearly shows that for a fixed value of the dopant density $N_D$, the internal electric field increases with decreasing film thickness $d$. Figure 4 also shows that for a fixed thickness $d$, the magnitude of the internal electric field increases with increasing dopant density $N_D$. As discussed in section 3, this behavior is indeed observed in the 3D simulations (Fig. 1); the internal electric field is the key to explain the mono-domain to multi-domain transition.

## 5. Influence of the Depletion Layers on Polarization Switching

Does the depletion layer play any role during the polarization switching process? Recently, Zubko *et al.* [16] studied the influence of depletion layers on hysteresis loops in thin films using a 1D Ginzburg-Landau model. In their monodomain study, they found that the depletion layers both tilt and shrink the hysteresis loops. This theoretical study [16] did not include effects associated with domain structure evolution during the switching process. We simulate the evolution of domain patterns under an applied external field by applying a time dependent (sawtooth with a $1.6 \times 10^6$ time step switching period) potential to each of the domain structures shown in Fig. 1. Before we describe our results on the polarization switching behavior, we remark that we are using a frozen depletion layer approximation (i.e., the depletion layer width does not change appreciably during polarization switching). This is a reasonable approximation for polarization switching that occurs quickly relative to any change in charge carrier profile. For example, domain wall velocities in $BaTiO_3$ can be in the

$10^{-7} - 10^5$ cm/s range, depending on the electric field [26-29]. This should be compared with the speed of oxygen vacancies; multiplying the oxygen vacancy drift mobility of $8.4 \times 10^{-22}$ m$^2$/(Vs) [20] by the intrinsic coercive field of BaTiO$_3$ (126 MV/m [30]) we find an oxygen vacancy velocity of $1.06 \times 10^{-13}$ m/s. Since the velocity of the oxygen vacancies is much, much smaller than domain wall velocities, it is appropriate to conclude that the depletion layer thickness does not significantly change during polarization switching at typical switching frequencies. Hence, the frozen depletion layer approximation should be valid here. On the other hand, if the mobile charges are primarily electrons and holes, or if we consider extremely low switching frequencies (of the order of the oxygen vacancy velocity, the results below should be viewed only as qualitative [31].

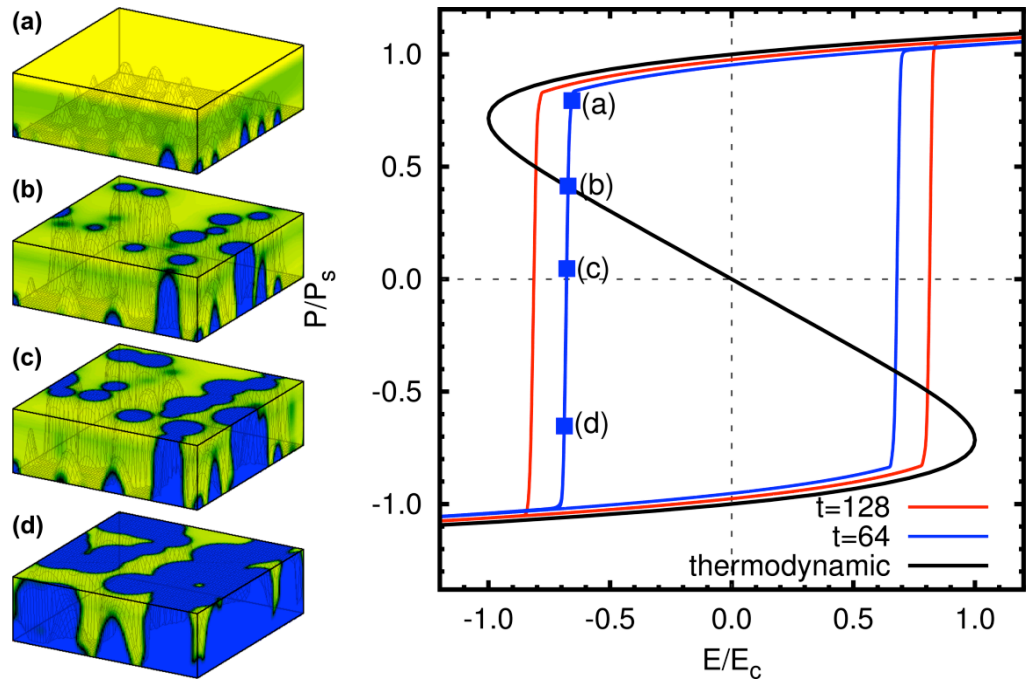

Figure 5 (color online). Hysteresis loops for $N_D = 4.11 \times 10^{25}$ m$^{-3}$, $d = 64, 128$nm. The domain patterns corresponding to points (a) to (d) for the $d = 64$ nm film are also shown. The video for nucleation and growth of domains through points (a) to (d) is provided in [32].

Figure 7 shows the simulated hysteresis loops for films with $d = 64$ and 128nm, along with the corresponding thermodynamic switching loop obtained from the homogeneous Landau theory (in the absence of the depletion layers). Note that the coercive field is lower than the thermodynamic coercive field, $E_c$, for both thicknesses. This is associated with the internal electric field $E_b$ which opposes the polarization and aids the switching process. Since $|E_b|$ increases with decreasing thickness (Fig. 6), the coercive field decreases with decreasing thickness. We also find that the switching process is inhomogeneous; it occurs via domain nucleation and growth. Images (a)-(d) in Fig. 7 depict this process for the $d$ = 64nm film. Nucleation of reverse domains occur where $E_z$ is most negative, i.e., at the bottom electrode ($z$ = 0) once $|E_z|_{z=0} > E_c$ as predicted by [2]. While many small reverse-domains nucleate during the switching process, the larger reverse-domains grow and nearby smaller reverse-domains to disappear. It is interesting to note that in our model the positions of the reverse-domains are random, and come from the thermal noise that we introduce into our model—we do not introduce any nuclei of reversed polarization or charged defects. The nucleated reverse-domains grow by lateral domain wall migration and coalescence. At sufficiently high electric field, the polarization in the entire film reverses. Thus, we observe that the depletion layers play a significant influence on the polarization switching process.

The shape of reverse domains formed during the switching process are similar to those observed in Fig. 1 under short circuit boundary conditions. To understand these domain shapes, it is instructive to examine the distributions of polarization and electric field in the vicinity of a reverse domain for a domain formed during the polarization switching process. Figure 8 shows the polarization vectors near the domain wall are approximately parallel to the wall. The electrostatic energy of the domain walls is large where the polarization vectors are head-to-head or tail-to-tail. Hence, these vectors rotate along the entire domain wall except at the tip of the reverse domain, where the electric field is the strongest (see Fig. 9). The rotated polarization vectors screen the electric field, and in the interior of the reverse domain, the electric field is effectively zero (Fig. 9). However, the magnitude of the electric field is approximately equal to the intrinsic coercive field, $E_c$, in the original domain immediately above the tip of the reverse domain.

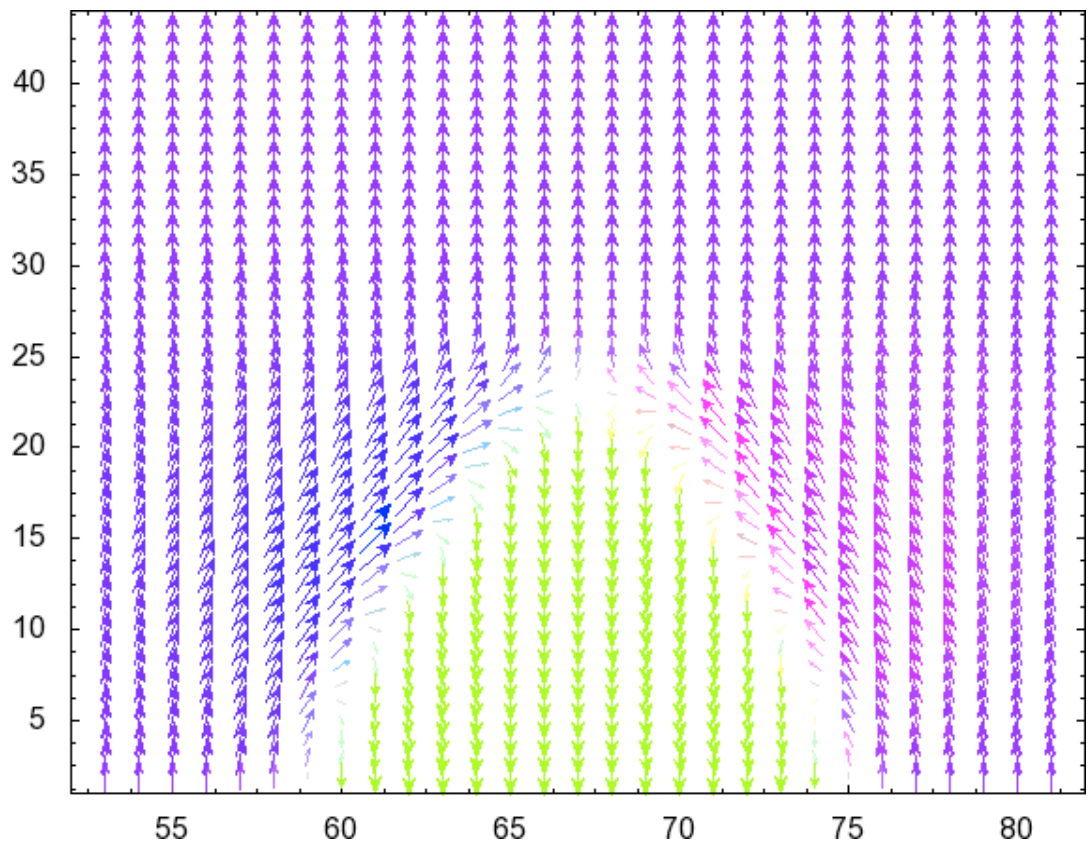

Figure 6 (color online): The polarization vectors in the vicinity of a conical reverse domain in the ferroelectric film near the lower contact. Note the rotation of the polarization vectors near the domain wall. Arrows are blue for $P_z > 0$ and yellow for $P_z < 0$.

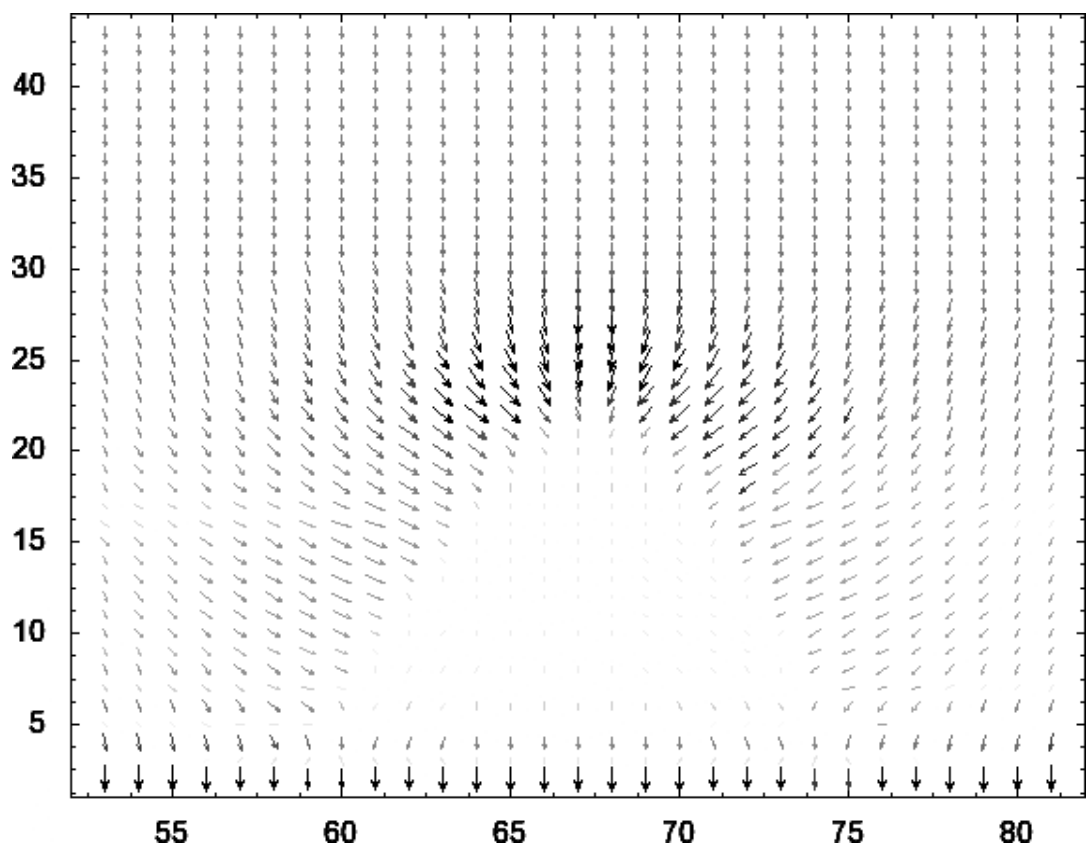

Figure 7: Electric field vectors corresponding to the domain wall configuration in Fig. 6. The electric field is strongest in the vicinity of the tip of the reverse domain (where polarization vectors are tail-to-tail) and weakest in the interior of the reverse domain.

## 6. Summary and Conclusion

In summary, using 3D phase field simulations we show that depletion layers at the ferroelectric thin film/electrode interface can create an internal electric field in the bulk regions where there is no free charge. This is also demonstrated explicitly through an analytical solution of a 1D Ginzburg-Landau model with depletion layers. We show that this internal electric field can have very important implications for thin film ferroelectrics. At high values of the donor dopant density and at thicknesses which are comparable to depletion layer width, a single domain remnant state can become unstable due to the appearance of a domain pattern. A multi-domain state with conical and cylindrical 180° domains (Fig. 1) is observed for such cases, resulting in a significant drop in the value of the remnant polarization. Further, it is found that the depletion layers also play an important role in domain nucleation during polarization switching. The phenomena reported in the present letter are distinct from those that are observed due to depolarization fields from uncompensated surface charges. While the depolarization fields due to bound charges become important only for very small thicknesses, the internal electric field due to depletion layers predicted by our calculations may become important for relatively thicker films, depending on the dopant density and the choice of the electrode.